\documentclass[reprint,onecolumn,showpacs,amsmath,amssymb,aps,superscriptaddress,floatfix,nofootinbib]{revtex4-2}
\usepackage{amsthm}
\usepackage{amsmath, mathtools}
\usepackage[utf8]{inputenc}	
\usepackage[english]{babel}
\usepackage{amsmath}
\usepackage{graphicx}		
\usepackage{natbib}
\usepackage{textcomp}
\usepackage{gensymb}
\usepackage[usenames,dvipsnames,svgnames,table]{xcolor}
\usepackage[hidelinks,colorlinks=false,urlcolor=Cerulean,citecolor=black]{hyperref}
\usepackage{siunitx}
\usepackage{mathrsfs}
\usepackage{multirow}
\usepackage{bm}
\usepackage{xfrac}
\usepackage{comment}
\usepackage[normalem]{ulem}
\usepackage{printlen}
\usepackage[T1]{fontenc}

\renewcommand{\i}{\mathrm{i}}

\begin{document}

\title{Phase-based spatial ordinal patterns for characterizing oscillatory dynamics}

\author{Robison J. Santos-Silva}
\affiliation{Department of Neuroscience, University of Lethbridge, Lethbridge AB, Canada}
\affiliation{Department of Physics, Federal University of Parana, Curitiba PR, Brazil}
\affiliation{Interdisciplinary Center for Science, Technology and Innovation (CICTI), Federal University of Parana, Curitiba PR, Brazil}

\author{Bruno R. R. Boaretto}
\affiliation{Institute of Science and Technology, Federal University of São Paulo (UNIFESP), São José dos Campos SP, Brazil}

\author{Thiago L. Prado}
\affiliation{Department of Physics, Federal University of Parana, Curitiba PR, Brazil}
\affiliation{Interdisciplinary Center for Science, Technology and Innovation (CICTI), Federal University of Parana, Curitiba PR, Brazil}

\author{Roberto C. Budzinski}
\email{roberto.budzinski@uleth.ca}
\affiliation{Department of Neuroscience, University of Lethbridge, Lethbridge AB, Canada}
\affiliation{Fields Lab for Network Computation, Fields Institute, Toronto ON, Canada}

\begin{abstract}
The emergence of organized spatiotemporal patterns is ubiquitous in oscillatory systems, from neural populations to engineered networks. Identifying these patterns and tracking how they evolve over time remains challenging, particularly when systems exhibit transient dynamics. Here, we introduce a framework based on spatial ordinal patterns to characterize the spatiotemporal dynamics of oscillatory systems. Our approach acts directly on the phase rather than the amplitude, with additional patterns introduced to account for near-equal phases. This symbolic representation encodes local spatial ordering relations, capturing both phase gradients and synchronized clusters within a single framework. From this construction, we define a spatial permutation entropy that quantifies the diversity of spatiotemporal patterns at each point in time, enabling the detection of transient dynamics and regime transitions as they occur. We show that this approach distinguishes phase-locked states with identical levels of global synchronization but distinct spatial organization and also characterizes partially synchronized states. We demonstrate the method on synthetic oscillator networks across multiple spatiotemporal regimes, and on resting-state EEG recordings from human volunteers, where it distinguishes different conditions within individual volunteers.
\end{abstract} 

\maketitle

\section*{Introduction}

Synchronization and pattern formation in oscillator networks are central phenomena in nonlinear dynamics, with applications ranging from physics and engineering to neuroscience and biology \cite{ermentrout2001traveling,motter2013spontaneous,rorato2017social,tyloo2019key,gu2025emergence}. In this context, networks of coupled oscillators have served as a central model for studying the emergence of synchronized patterns \cite{strogatz2000kuramoto,arenas2008synchronization,rodrigues2016kuramoto}. Depending on the interplay between intrinsic frequencies, coupling architecture, and phase lags, oscillator networks display a rich repertoire of collective behaviours, including phase synchronization \cite{rosenblum1996phase,wiley2006size,delabays2017size}, clustering and chimera states \cite{abrams2004chimera,omel2018mathematics}, and travelling waves \cite{jeong2002time,ko2007effects,laing2016travelling,budzinski2023analytical}. A central challenge in the study of such systems is to characterize how macroscopic organization emerges from network interactions, particularly during transient dynamics and transitions between dynamical regimes. While mean-field quantities provide a powerful characterization of global coherence, capturing the full mesoscopic structure of these systems -- including partially synchronized states, chimera configurations, and distinct phase-locking solutions -- requires observables that retain spatial information while remaining statistically tractable.

Ordinal analysis offers a natural framework for such a description \cite{bandt2002permutation}. Originally introduced for time-series analysis, ordinal patterns characterize local ordering relations among neighbouring values, providing a symbolic representation of dynamics that is robust to noise \cite{bandt2002permutation,leyva202220}. Symbolic analysis and the associated permutation entropy have since been applied widely, including to chaotic and stochastic time series \cite{rosso2007distinguishing,politi2017quantifying,boaretto2021discriminating,zanin2021ordinal}, neural dynamics \cite{aragoneses2014unveiling,masoliver2018sub,budzinski2021symbolic}, laser dynamics \cite{soriano2011time,toomey2014mapping,gancio2025identifying}, biomedical signals \cite{zanin2012permutation,parlitz2012classifying}, and inference of network connectivity \cite{bahraminasab2008direction,tirabassi2015inferring,zhang2017constructing}, among others. It is also possible to extend ordinal analysis from the temporal to the spatial domain: rather than ordering values along time, spatial ordinal patterns order values across neighbouring locations at a fixed time point. This approach is particularly suited to systems with both spatial and temporal organization, such as networks, and has been used to identify distinct brain states \cite{boaretto2023spatial} and to perform image classification \cite{Ribeiro2012image,zunino2016discriminating,bandt2023two}.

Here, we extend the spatial ordinal framework to characterize the spatiotemporal organization of oscillatory dynamics. Two extensions are central to this approach. First, while ordinal patterns have so far been applied to amplitude-like variables, we adapt the method to the circular nature of oscillatory phase, and introduce additional patterns that explicitly account for near-equal phases. The resulting symbolic representation captures both local phase gradients and synchronized clusters within a single description. Second, building on this representation, we define a spatial permutation entropy that quantifies the diversity of local phase configurations at each time step, enabling the study of transient dynamics as they unfold. This measure provides a compact yet structurally informative description of the system's state, allowing us to track regime transitions, identify clustering phenomena, and distinguish between phase-locked solutions that share the same order parameter but differ in spatial structure. We demonstrate this approach on synthetic oscillator networks across a range of spatiotemporal regimes, and apply it to resting-state EEG recordings (obtained from open-dataset in \cite{PhysioNet-eegmmidb-1.0.0}), where our approach distinguishes eyes-open from eyes-closed conditions within individual participants. Together, these results show that spatial ordinal patterns reveal dynamical structure in oscillator networks and empirical data at the meso- and macroscopic level, including transient regimes and individual cases.

\section*{Ordinal patterns, spatial permutation entropy, and spatiotemporal dynamics}

\begin{figure}[htb]
    \centering
    \includegraphics[width=0.95\linewidth]{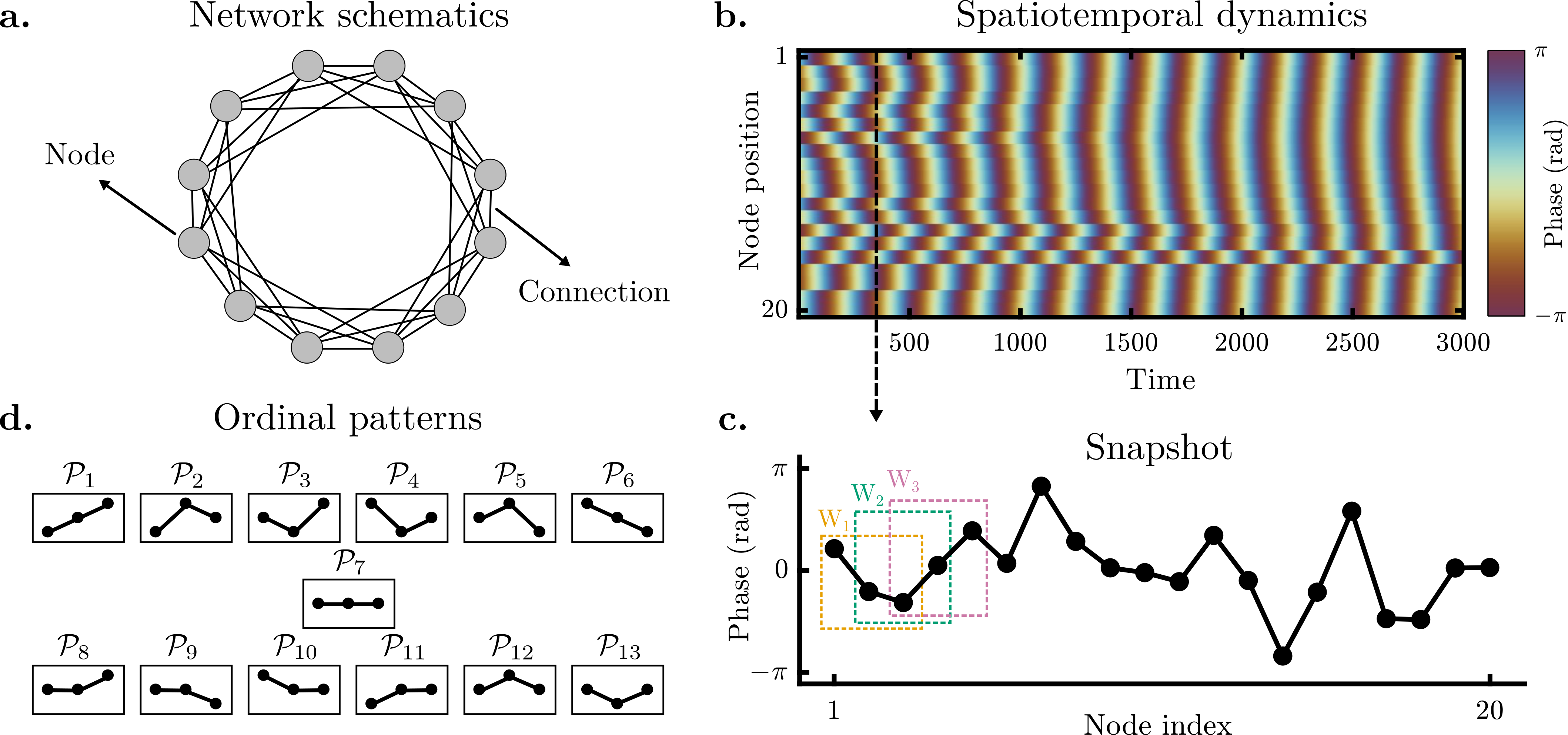}
    \caption{\textbf{Spatial ordinal analysis on oscillator networks.} \textbf{(a)} Schematic representation of a network, where each grey dot represents a node or an oscillator and the black lines represent the connections. \textbf{(b)} Example of spatiotemporal dynamics in oscillator networks. Here the phase is plotted in colour-code, with time in the horizontal axis and node position in the vertical axis. As time evolves, the phase pattern changes due to the interactions in the network. \textbf{(c)} Snapshot of phase pattern for a specific time. We analyze the phase pattern of consecutive nodes with overlapping windows using circular distance. \textbf{(d)} We characterize the phase patterns with these 13 ordinal patterns. The patterns with equal phase are analyzed under a specific threshold $\sigma$. After obtaining the probability of each pattern, we then evaluate the permutation entropy, given by Eq.~\eqref{eq:entropy}. This procedure is performed for each timestep.}
    \label{fig:schematics_method}
\end{figure}

In this paper, we use ordinal patterns and spatial permutation entropy to study the phase dynamics of oscillatory systems. To introduce the approach and build up intuition, we can start with a simple example of an network of oscillators (Fig.~\ref{fig:schematics_method}a). Due to the interactions in the system, the dynamics of the network unfolds over space and time -- see Methods Sec.~\ref{sec:kuramoto} for details on the model and simulations. We can observe this evolution on the phase dynamics plotted in colour-code as a function of node position and time (Fig.~\ref{fig:schematics_method}b). In this example, the network started at a random, asynchronous state, but as time evolves, we can appreciate the emergence of organization in the system. Our goal is to study how these patterns evolve over space and time. To do so, we consider snapshots of the phase dynamics of all oscillators for each instant of time (Fig.~\ref{fig:schematics_method}c). With this, we can divide the phase patterns into overlapping windows ($W_1, W_2, W_3$, Fig.~\ref{fig:schematics_method}c), where we then use the ordinal patterns to characterize the dynamics (Fig.~\ref{fig:schematics_method}d).

Ordinal analysis is a popular tool to identify bifurcation and characterize time series. Here, we study ordinal patterns using overlapping windows of size 3 on the phase values over node position. In addition to the traditional 6 patterns ($\mathcal{P}_1, \cdots, \mathcal{P}_6$), we also consider 7 new patterns that consider (partially) synchronized nodes ($\mathcal{P}_7, \cdots, \mathcal{P}_{13}$, Fig.~\ref{fig:schematics_method}d). The reason behind this choice streams from the fact that oscillator system have been shown to display clusters of synchronized nodes in the transition to synchronization \cite{bayani2024transition}, which may have an impact in neural function \cite{bansal2019cognitive,masoliver2025hippocampal} and computation \cite{budzinski2024exact}. These additional patterns enhance our ability to distinguish fine details in the transitions between states (see Supplement, Figs.~S1 and S2).

These 13 patterns give a complete description of all possible states of size 3 that can exist in the network. Because we are studying phase dynamics, we use circular distance between phase values. Further, due to float point precision and heterogeneity in the system, the condition where two phase values are considered equal in our approach depends on a parameter $\sigma$. In practice, this a free parameter that can be adjusted for each analysis. Importantly, however, $\sigma$ can have a physical interpretation. For example, in the case of synchronization of oscillators, the $\sigma$ represents the maximum phase difference between two oscillators to be considered ``in-sync''. We note that different values of $\sigma$ are possible, and while this has to be tuned for each application, the results are robust to a range of parameters, instead of being sensitive to very specific values (see Supplement, Fig.~S3).

For a given snapshot of the phase dynamics, we can then obtain the probability of occurrence of each pattern $\mathcal{P}$, for each instant of time of the dynamics evolution. We then evaluate the permutation entropy
\begin{equation}
    S(t) = - \sum_{j=1}^M P\big(\mathcal{P}_j(t)\big) \ln{\Big( P\big(\mathcal{P}_j{(t)\big) \Big)}},
    \label{eq:entropy}
\end{equation}
where $P\big(\mathcal{P}_j(t)\big)$ is the normalized probability of the pattern $\mathcal{P}_j$ at time $t$, and $M = 13$ is the total number of patterns. We can then normalize the entropy $S$ by its maximum possible value -- when the probability of all patterns is equal -- which gives us
\begin{equation}
    H(t) = \frac{S(t)}{\ln{(M)}},
    \label{eq:normalized_entropy}
\end{equation}
so that $H(t) \in [0, 1]$, with $H = 1$ corresponding to the case where all patterns are equally probable, and $H = 0$ to a single repeated pattern. Repeating this procedure at every timestep yields a time-resolved signal that tracks the evolution of spatial organization throughout the dynamics.

\section*{Transition to synchronization in oscillator networks}

\begin{figure}[htb]
    \centering
    \includegraphics[width=0.8\linewidth]{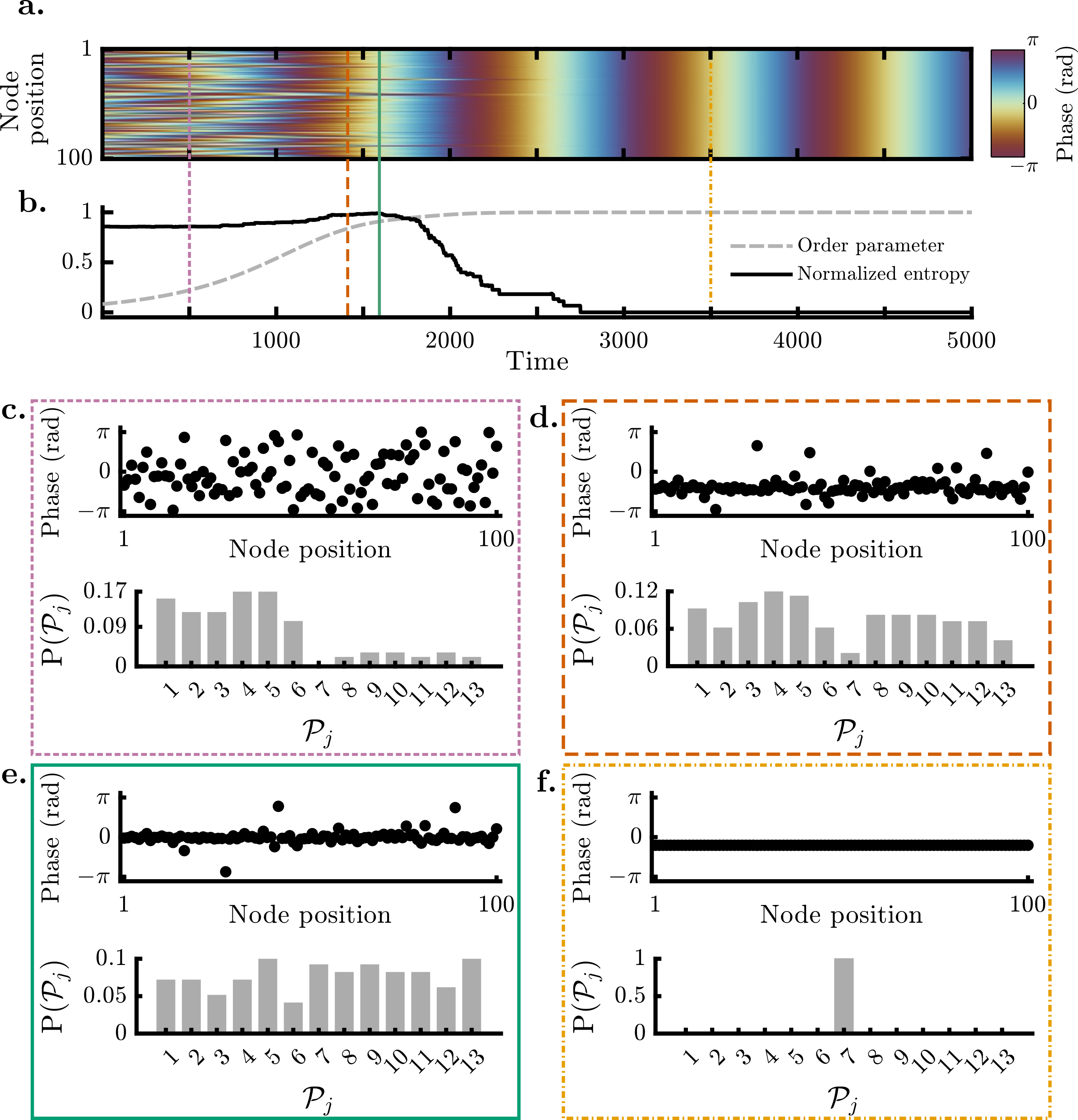}
    \caption{\textbf{Transition to phase synchronization.} We consider a globally connected network of Kuramoto oscillators (see Methods Sec.~\ref{sec:kuramoto}), with $N = 100$ nodes, natural frequency $\omega_i = 0, \, \forall \, i$, $\epsilon = 0.04$, phase-lag $\phi_{ij} = 0, \, \forall \, ij$. \textbf{(a)} We start the network with random initial phases, but due to the coupling, the network transitions to the phase synchronized state. \textbf{(b)} The order parameter (dashed grey line) starts at a residual value due to finite size effects, and it grows to one as the network transitions to phase synchronization. The normalized permutation entropy (black line) is high ($H \approx 0.8$) when the network is asynchronous in the beginning. As the network starts to transition to phase synchronization, the permutation entropy grows at the transition ($H \approx 1$), and as the network synchronizes, the permutation entropy goes to zero. \textbf{(c)} We consider exemplary snapshots of the phase pattern at different times and also analyze the probability of different ordinal patterns. When the network is asynchronous (pink dashed line), patterns 1-6 are almost equally probable, but the other patterns are less probable. \textbf{(d-e)} During the transition (dashed orange and solid green lines), the network starts to display clusters of synchronized nodes, which increases the probability of occurrence of patterns 7--13. Because of that, the entropy peaks. \textbf{(f)} As the network reaches the phase synchronized state (dotted-dashed yellow line), only pattern 7 is present -- because all the phase are equal -- and thus the entropy is equal zero. For the analyses here, we consider the threshold $\sigma = \sfrac{\pi}{30}$.}
    \label{fig:transition_sync_example}
\end{figure}

As a first application of this framework, we study the traditional transition to phase synchronization in oscillator networks (details on the model and simulation are in Methods Sec.~\ref{sec:kuramoto}). We start by considering the case where all oscillators have the same natural frequency and are connected to all other nodes in the network, which leads to the phase synchronized state being the only stable solution for the system. In fact, the network starts in an asynchronous state due to the random initial conditions, but transitions to a phase synchronized state (Fig.~\ref{fig:transition_sync_example}a). To characterize this transition, we first calculate the Kuramoto order parameter $R$ as a function of time (see Methods Sec.~\ref{sec:order_parameter}), a traditional quantification of phase synchronization, where $R = 1$ indicates phase synchronization, and random distributed phases scale with $R \sim \sfrac{1}{\sqrt{N}}$, where $N$ is the number of oscillators in the system. The order parameter starts at low values, and as the network transitions to phase synchronization, the quantifier reaches one (dashed-grey line, Fig.~\ref{fig:transition_sync_example}b).

We then use the spatial ordinal analysis and the permutation entropy Eq.~\eqref{eq:normalized_entropy} to study this transition. We calculate the ordinal patterns probabilities for each timestep of the dynamics evolution and then obtain the permutation entropy (black line, Fig.~\ref{fig:transition_sync_example}b), which displays values $H \approx 0.8$ when the system is desynchronized in the beginning of the simulation. During the transition, however, the entropy increases and reaches $H \approx 1$, and as the network reaches a phase synchronized state, the entropy vanishes and reaches $H = 0$.

To gain insights into this process, we analyze snapshots of the phase dynamics and the probabilities of the ordinal patterns at different times in the evolution of the dynamics. When the network is in an asynchronous state (dashed pink line, Fig.~\ref{fig:transition_sync_example}b), the phase dynamics is similar to a random pattern, and the probabilities of the ordinal patterns show an almost probabilities for the first 6 patterns and low for the remaining 7 (Fig.~\ref{fig:transition_sync_example}c). During the transition to synchronization (dashed orange and solid green lines, respectively, Fig.~\ref{fig:transition_sync_example}b), the network displays clusters of synchronized nodes, and with that, the probabilities of ordinals patterns 7-13 increase, reaching a state where all patterns 1--13 are almost equally probably (Figs.~\ref{fig:transition_sync_example}d and \ref{fig:transition_sync_example}e), which in turns leads to the maximum ordinal entropy. Finally, when the network is in a phase synchronized state (dotted-dashed yellow line, Fig.~\ref{fig:transition_sync_example}b), only pattern 7 is present (Fig.~\ref{fig:transition_sync_example}f), and thus $H = 0$. Together, these results suggest that the transition to synchronization in this network is mediated by the emergence of synchronized clusters, and that our framework is able to capture these details with the additional, partially synchronized ordinal patterns. 

\section*{Characterizing different dynamical states}

\begin{figure}
    \centering
    \includegraphics[width=0.865\linewidth]{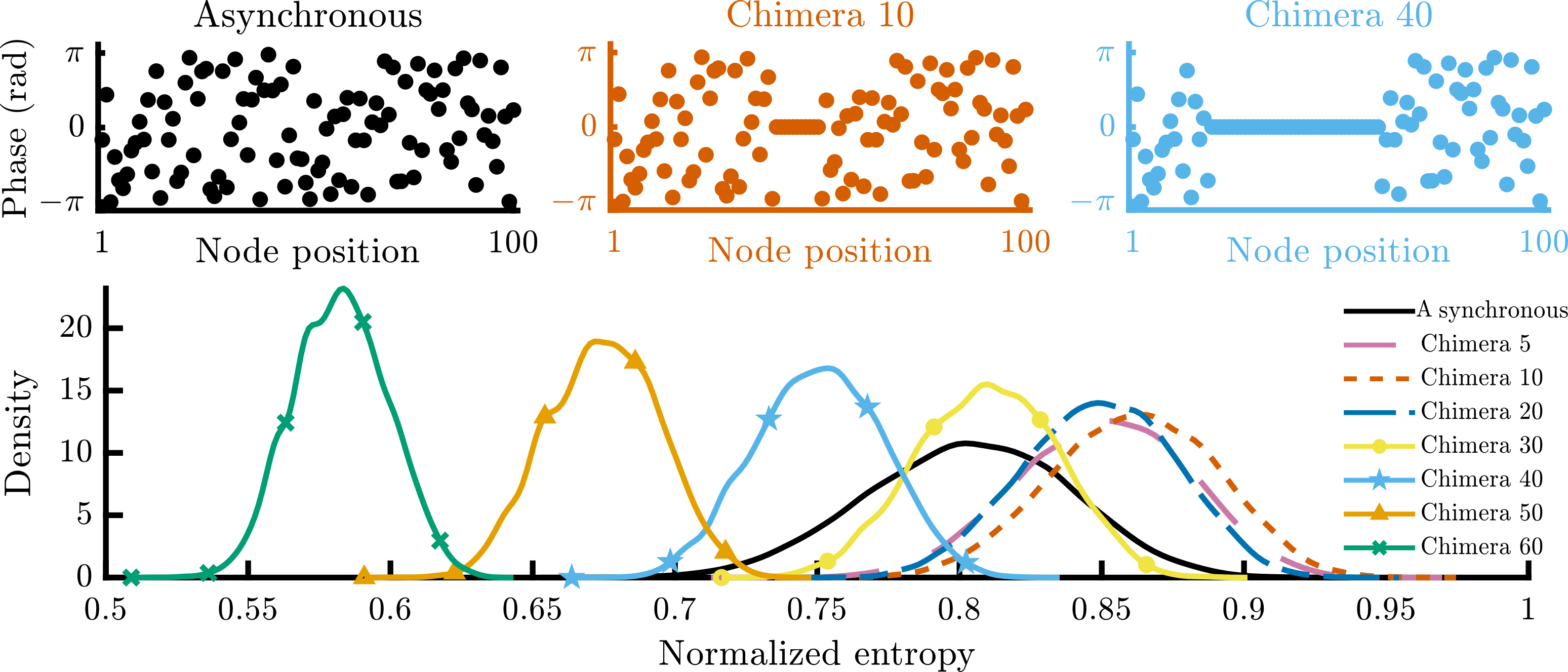}
    \caption{\textbf{Spatial permutation entropy of partially synchronized states.} We consider different phase patterns that are relevant for oscillator networks and calculate the entropy. First, we consider $100,000$ different realizations of random, asynchronous patterns (black line), where the entropy varies from 0.7 to 0.9, with mean around 0.82. We also consider chimera states with different sizes of the synchronized cluster. The cases where the synchronized clusters vary from 5\% to 30\% have higher entropy than the random asynchronous case. As the synchronized cluster grows in size, the entropy decreases, because pattern 7 dominates. For the analyses here, we consider the threshold $\sigma = \sfrac{\pi}{30}$.}
    \label{fig:chimeras}
\end{figure}

This framework, in fact, allows us to identify a diversity of patterns that are relevant in networked systems. To explore this in more details, we consider simulations of oscillator networks (see Methods Sec.~\ref{sec:kuramoto}) that lead to a variety of different states. We first consider asynchronous states, where the phases are random (example on top left, Fig.~\ref{fig:chimeras}). We also consider states where part of the network is synchronized, while the other part remains asynchronous, where the size of the synchronized cluster can vary (examples on top middle and top right, Fig.~\ref{fig:chimeras}). These states are known as ``chimeras'', and were first observed in networks of coupled oscillators \cite{abrams2004chimera}. 

\begin{figure}[b!]
    \centering
    \includegraphics[width=0.875\linewidth]{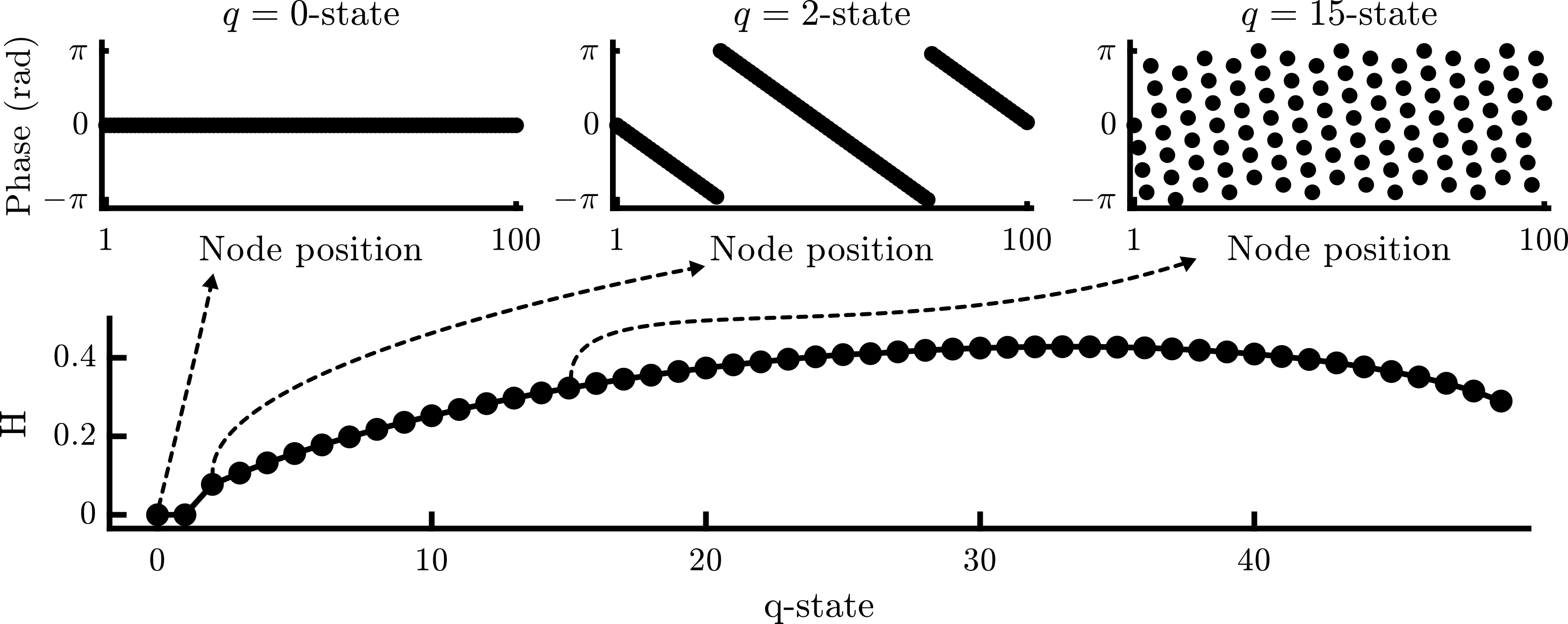}
    \caption{\textbf{Spatial permutation entropy of phase-locked states.} We study the spatial permutation entropy of phase-locked states, or q-states, which are solutions given by: $\theta_{j}^{(q)} = \frac{2\pi q}{N}j$. In this case, $q = 0$ representing the phase synchronized state, and $|q|>0$ representing waves in the network. For patterns with low-spatial frequency (low $|q|$), the entropy is low, but as the spatial frequency increases, the entropy also increases. However, as the spatial frequency becomes very high ($|q| > 34$), the entropy decreases again, highlighting the non-monotonic nature of this process. We note these states can be obtained in Kuramoto networks with distance-dependent delays. For the analyses here, we consider the threshold $\sigma = \sfrac{\pi}{30}$.}
    \label{fig:q-states}
\end{figure}
We then use the spatial permutation entropy and the framework introduced here to analyze these states. Specifically, we consider $100,000$ different realizations of each of these patterns, where we calculate the spatial entropy of each of them. Figure \ref{fig:chimeras} (bottom) shows the distribution of normalized entropy $H$ for different patterns over all the realizations. First, for the random, asynchronous state, the mean of the distribution is centred in $H \approx 0.82$ with a big dispersion due to the finite nature of the network (black line, Fig.~\ref{fig:chimeras}). To systematically study the entropy of chimera states, we consider states where the synchronized cluster's size varies from 5 to 60 nodes. We observe that, when the synchronized cluster is between 5\% and 30\% (dashed orange, dashed blue, dashed pink, and solid yellow lines, Fig.~\ref{fig:chimeras}), the entropy is higher than the random asynchronous state (black line, Fig.~\ref{fig:chimeras}). As the synchronize cluster grows in size, the entropy decreases, as pattern 7 becomes dominant (solid blue, solid orange, and solid green lines, Fig.~\ref{fig:chimeras}).

We also use this framework to study different synchronized states. Specifically, we consider phase-locked or $q$-states, which are often equilibria for the Kuramoto model: $\bm{\theta}^{(q)} = \big(0, \frac{2\pi q}{N}, \cdots, \frac{2\pi q(N-1)}{N} \big)$. In this case, $q = 0$ represents the phase-synchronized state (top left, Fig.~\ref{fig:q-states}), where all oscillators have the same phase and frequency, and the other solutions represent phase-locked states or $q$-twisted states, where the oscillators have a constant phase offset across nodes (top middle and right, Fig.~\ref{fig:q-states}). These states represent waves travelling around the network with different spatial frequencies. These solutions can be observe in a diversity of networks, including systems with distance-dependent delays. Specifically, different combinations of connectivity and delays can stabilize $q$-states of different spatial frequency \cite{sinha2025geometric}. We then study the normalized entropy $H$ of the entire range of $q$-states for a network with $N = 100$ nodes. 

Our results show that for the phase-synchronized state $q=0$ and for the $|q| = 1$ state, the entropy is zero, but for the other $q$-state, the entropy is non-zero (bottom, Fig.~\ref{fig:q-states}) -- considering the threshold $\sigma = \sfrac{\pi}{30}$. As the spatial frequency increases (represented by higher values of $|q|$), the normalized entropy $H$ also increases, which indicates that different ordinal patterns are contained in the state. However, this behaviour is non-monotonic, and for higher spatial frequency states ($|q| > 34$), the spatial permutation entropy decreases again.

\section*{Different paths to synchronization various networks}

\begin{figure}[htb]
    \centering  
    \includegraphics[width=0.85\linewidth]{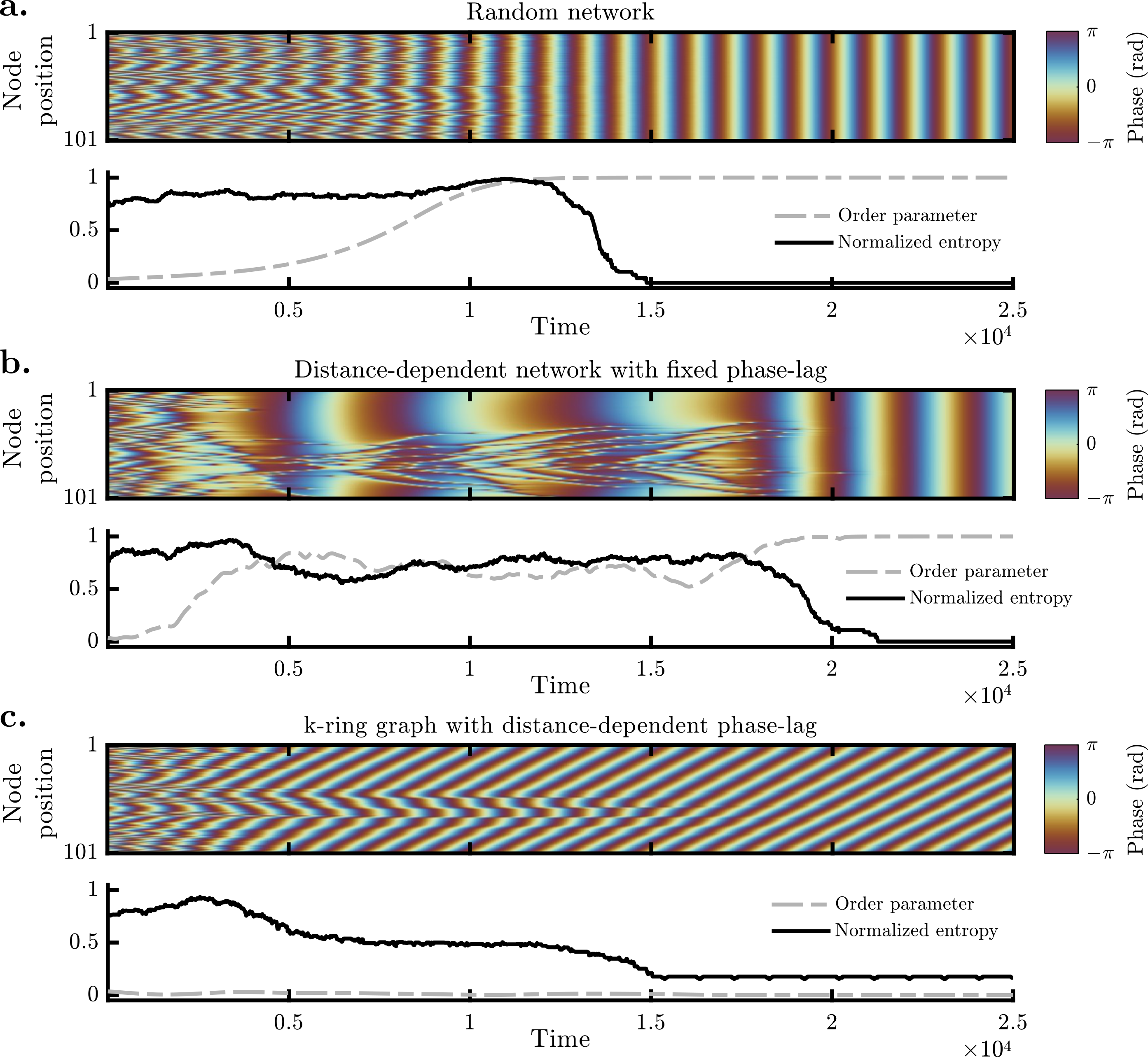}
    \caption{\textbf{Transition to synchronization in different networks.} \textbf{(a)} We first consider a random network of oscillators. The network starts with asynchronous dynamics because of the random initial state, but due to the coupling, the network transitions to phase synchronization. This is reflected by the order parameter (dashed grey line) and the non-monotonic behaviour of the normalized entropy (black line). \textbf{(b)} We also consider a distance-dependent network with fixed phase-lag. In this case, the network transitions from the initial random condition to display a transient chimera before transitioning to phase synchronization. In this case, the order parameter (dashed grey line) and the normalized entropy (black line) assume intermediate values due to the rich dynamics, before the network reaches phase synchronization. \textbf{(c)} We also consider a k-ring graph with distance-dependent phase-lags. In this case, phase synchronization is no longer a stable solution, but a different q-state instead. Here, we observe the random initial dynamics, transitioning to a mix of q-states, before the network reaches a stable q-state. The order parameter (dashed grey line) remains low and assumes zero when the stable state is reached. The normalized entropy (black line) highlights the details on the transition, showing specific plateaus where different dynamics happen. For the analyses here, we consider the threshold $\sigma = \sfrac{\pi}{30}$.}
    \label{fig:transitions}
\end{figure}

We further study the transition to synchronization under different conditions. First, we consider a oscillator network on a random graph (see Methods Sec.~\ref{sec:kuramoto} for details). In this case, the phase synchronized state is a solution. However, the transition over time to this solution differs from the example in Fig.~\ref{fig:transition_sync_example}. Specifically, the transition takes longer, which a richer diversity of patterns before the final synchronized state is reached (top Fig.~\ref{fig:transitions}a). The order parameter shows this transition (dashed grey line, bottom Fig.~\ref{fig:transitions}a), and the normalized entropy $H$ reveals fine variations in the patterns during this transition, with peak just before the final synchronized state is reached and $H = 0$ (black line, bottom Fig.~\ref{fig:transitions}a).

We also consider a network with distance-dependent coupling and a fixed, homogeneous phase-lag in the coupling (see Methods Sec.~\ref{sec:kuramoto}). While phase synchronization is still a stable solution, the transient can be indeed long, where chimera states appear \cite{wolfrum2011chimera}. We start the network at a random initial state, but due to the interactions in the network, the system displays states with coherent clusters and asynchronous parts -- a chimera state (top Fig.~\ref{fig:transitions}b). After a the transient, where chimeras are observed, the system reaches a phase synchronized state. When we calculate the order parameter and spatial permutation entropy (dashed grey and black lines, respectively, bottom Fig.~\ref{fig:transitions}b), we can appreciate the complexity of the transient dynamics. Further, together with the results discussed in Fig.~\ref{fig:chimeras}, we can interpret the specific values of $H$ and its variation over time directly in terms of the size of the synchronized cluster and how it varies, which offers a new lens on this transient state.

Lastly, we study a network with nonlocal coupling (k-ring graph) and distance-dependent phase-lags, which mimic delays in the network. The combination of connectivity and delays can destabilize the phase synchronization, and instead, lead to a different phase-locking solution \cite{budzinski2023analytical,sinha2025geometric}, the $q$-states. In this case, the network starts with random initial conditions and then evolves to a series of states that represent waves travelling on the network, which are represented by the diagonal lines in the spatiotemporal dynamics (top Fig.~\ref{fig:transitions}c). We then evaluated the order parameter, which remains low for then entire simulation (dashed grey line, bottom Fig.~\ref{fig:transitions}c). This is because $R \sim \sfrac{1}{\sqrt{N}}$ for random, asynchronous phases, and $R = 0$ when the dynamics is given by a $q$-state (with $q \neq 0$). When we calculate the normalized entropy $H$, however, we can see the richness of the dynamics and the different transitions that happen (black line, bottom Fig.~\ref{fig:transitions}c). The normalized entropy starts at $H \approx 0.8$ due to the random initial conditions and just before the network starts the transition to the wave states, the normalized entropy displays a peak. When the network shows a complex combination of waves, the normalized entropy assumes $H \approx 0.5$, and when the network reaches its asymptotic state ($q=6$-state), the normalized entropy reaches a state $H \approx 0.17$, characteristic of that state (see Fig.~\ref{fig:q-states}).

Taken together, these results show the framework introduced here allows us to characterize different patterns in oscillator networks and how they evolve over time in a variety of systems. Importantly, the spatial permutation entropy based on the symbolic analysis with 13 patterns (including the near-sync patterns) is able to capture traditional transitions between dynamical state and also reveal fine details between them.

\section*{Characterizing brain resting states in EEG signals}

\begin{figure}[htb]
    \centering
    \includegraphics[width=0.95\linewidth]{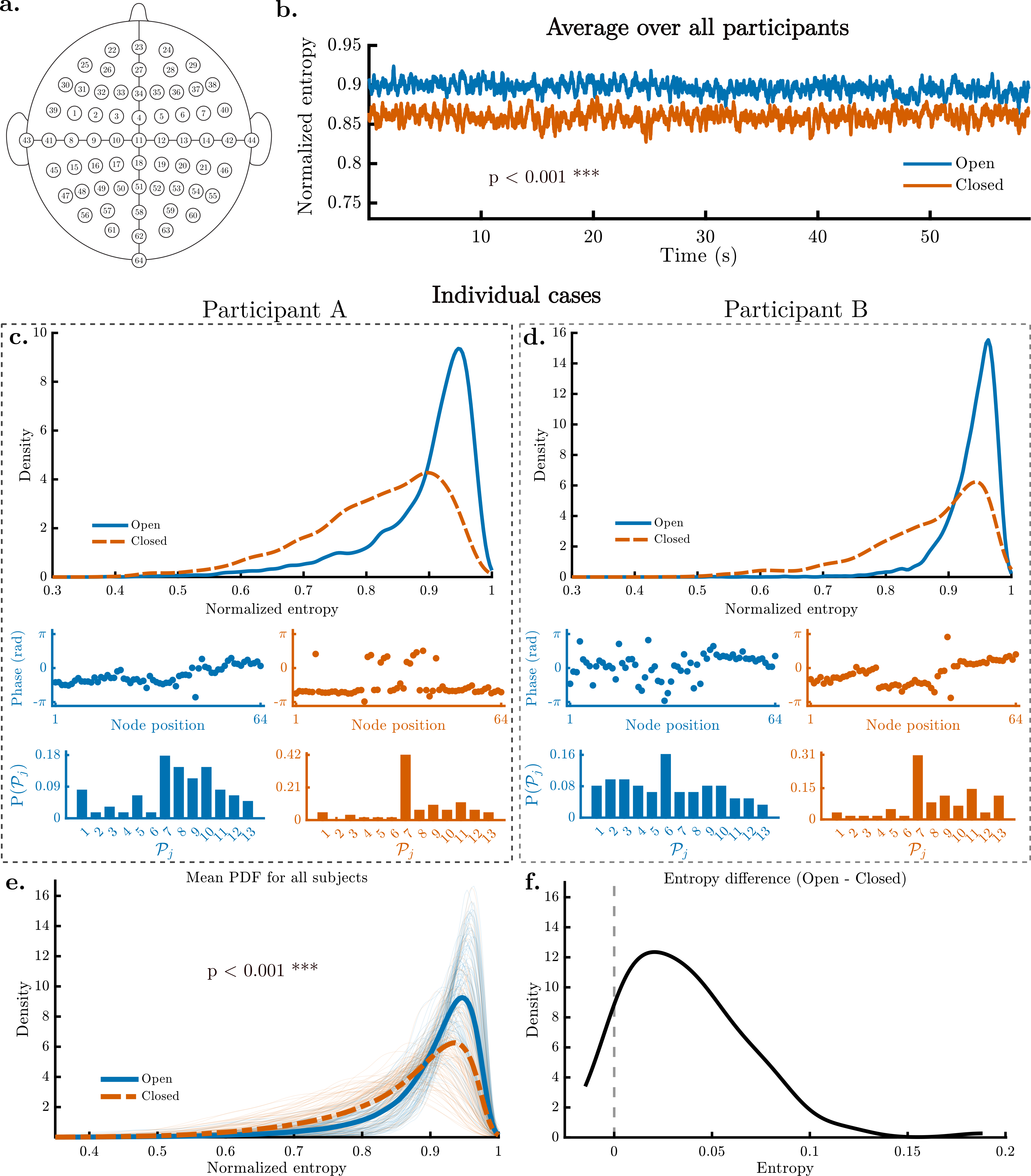}
    \caption{\textbf{Phase-based spatial ordinal analysis distinguishes resting-state brain activity.} \textbf{(a)} Resting-state EEG signals from 64 scalp channels are filtered in the $\alpha$-band (8--12 Hz) and the instantaneous phase is extracted. \textbf{(b)} Spatial permutation entropy $H(t)$ computed from the phase dynamics separates eyes-open from eyes-closed conditions when averaged across all volunteers. \textbf{(c-d)} The framework applied to two representative individual participants. Top: $H(t)$ reliably distinguishes the two conditions in both cases, here represented by the density of the normalized entropy. Bottom: Pattern probability distributions show that eyes-open recordings yield broadly distributed patterns and higher entropy, whereas eyes-closed recordings are dominated by pattern 7, reflecting increased spatial synchronization and lower entropy. \textbf{(e-f)} Across all 109 volunteers, the eyes-open state exhibits systematically higher entropy than eyes-closed in 104 cases. For the analyses here, we consider the threshold $\sigma = \sfrac{\pi}{18}$.}
    \label{fig:eeg}
\end{figure}

We now apply our framework to empirical data. We consider resting-state electroencephalography (EEG) recordings from healthy human volunteers \cite{PhysioNet-eegmmidb-1.0.0,schalk2004bci2000}, comprising signals from 64 channels distributed across the scalp (Fig.~\ref{fig:eeg}a). The dataset includes 109 volunteers, each recorded for two minutes, with one minute under the eyes-open condition and one minute under the eyes-closed condition. Distinguishing brain activity between these states is an important problem in neuroscience, and different methods have been proposed for this purpose \cite{barry2007eeg,petro2022eyes}, including ordinal and symbolic analyses applied directly to the EEG amplitude signal \cite{boaretto2023spatial,gancio2024permutation}.

Rather than working with the raw amplitude, we take a different route and extract the phase from the EEG signal filtered in the $\alpha$-band (8--12 Hz), which is known to play a central role in distinguishing eyes-open from eyes-closed activity \cite{barry2007eeg}. This yields phase dynamics across 64 channels for each participant and condition (see Methods, Sec.~\ref{sec:eeg_data}). We then apply our framework directly to these phase dynamics, obtaining the probability of each of the 13 ordinal patterns and the spatial permutation entropy $H(t)$ as a function of time.

Averaged across all participants, $H(t)$ clearly separates the two conditions (Fig.~\ref{fig:eeg}b), consistent with previous ordinal analyses of this dataset \cite{boaretto2023spatial}. Crucially, our approach extends beyond group-level averages and enables discrimination at the level of individual volunteers. For two representative participants (A and B), the framework reliably distinguishes eyes-open from eyes-closed conditions (Figs.~\ref{fig:eeg}c--d, top). During eyes-open recordings, pattern probabilities are broadly distributed across all 13 patterns, resulting in higher entropy; during eyes-closed recordings, pattern 7 becomes dominant, reflecting increased spatial synchronization and yielding a correspondingly lower entropy (Figs.~\ref{fig:eeg}c--d, bottom). This individual-level discrimination holds broadly across the dataset (Fig.~\ref{fig:eeg}e): the eyes-open state exhibits systematically higher entropy than the eyes-closed state in 104 of 109 volunteers (Fig.~\ref{fig:eeg}f). Lastly, we remark these results depend on the spatial relationships among electrodes rather than their individual signals: the discrimination is preserved under an alternative electrode arrangement \cite{boaretto2023spatial} (Supplemental Fig.~S4), but the distinction power is lost when electrode positions are randomly shuffled (Supplemental Fig.~S5).

\section*{Discussion and conclusion}

We have introduced a spatial ordinal framework to characterize the moment-by-moment spatiotemporal dynamics of oscillatory systems. By adapting ordinal pattern analysis to circular phase variables and incorporating additional patterns that account for near-equal phases, the method captures local spatial structure while remaining computationally tractable. The resulting symbolic representation encodes phase gradients, synchronized clusters, and phase-locking configurations within a single description, allowing us to obtain a fine description of transitions between different patterns. The associated spatial permutation entropy provides a time-resolved quantification of a diversity of configurations, enabling the detection of transient regimes and dynamical transitions as they occur. A key result is that states with identical levels of global synchronization can exhibit distinct ordinal signatures, demonstrating that the method accesses mesoscopic spatial structure that mean-field observables might miss. This discriminatory power is not limited to a single dynamical regime: across a range of spatiotemporal patterns -- from synchronization transitions to chimera states and travelling waves -- the spatial permutation entropy tracks the evolution of collective organization where the ordinal patterns can reveal fine details on the dynamical states.

Spatiotemporal organization and the interactions that give rise to it are central to the function of a broad class of systems, from abstract oscillator networks, to artificial neural networks and biological neural circuits \cite{muller2018cortical,csaba2020coupled,ricci2021kuranet,budzinski2024exact,effenberger2025functional}. In the brain in particular, the spatial patterning of oscillatory activity has been increasingly recognized as a carrier of functional information \cite{palmigiano2017flexible,breakspear2017dynamic,fries2023rhythmic}. The phase dynamics of neural populations encode spatial relationships that may support function and behaviour \cite{davis2020spontaneous,mohan2024direction}. Our framework is naturally suited to this setting, operating directly on phase rather than amplitude and preserving the spatial relationships among recording sites that purely temporal analyses discard.

This is reflected in our EEG application, where the framework distinguishes eyes-open from eyes-closed resting states at the level of individual volunteers. The eyes-closed state is characterized by the dominance of a single synchronized pattern and lower entropy, consistent with the well-established enhancement of $\alpha$-band synchronization under eyes-closed conditions \cite{hohaia2022occipital}. The eyes-open state, by contrast, yields broadly distributed pattern probabilities and higher entropy, reflecting greater spatial heterogeneity in phase organization. That this discrimination is lost when electrode positions are randomly shuffled confirms that the method captures genuinely spatial structure rather than a repackaging of individual channel signals.

Taken together, these results establish spatial ordinal analysis as a general and interpretable tool for probing mesoscopic organization in oscillatory systems. Because the framework is model-independent and relies only on phase ordering relations, it applies readily to oscillatory systems with heterogeneous structure, delayed coupling, or complex synchronization transitions -- in both synthetic and empirical settings. The symbolic representation is lightweight enough to scale to large networks, yet sensitive enough to resolve distinctions that global measures miss. We anticipate that this approach will find application wherever the spatial patterning of oscillatory phase dynamics carries information about the system's organization or function, including neural recordings, coupled oscillator models, and spatially extended dynamical systems more broadly.

\section*{Methods}

\subsection{Oscillator networks and simulations}\label{sec:kuramoto}

To simulate spatiotemporal dynamics in oscillator networks, we consider the Kuramoto model \cite{kuramoto1984cooperative,acebron2005kuramoto,rodrigues2016kuramoto}, which is given by:
\begin{equation}
    \dot{\theta}_{i}(t) = \omega_{i} + \epsilon \sum_{j=1}^{N} A_{ij} \sin{\Big(\theta_j(t) - \theta_i(t) - \phi_{ij}\Big)},
    \label{eq:km_model}
\end{equation}
where $\theta_i(t) \in [\pi, \pi)$ is the phase of the oscillator $i$ at time $t$, $\omega_i$ is the natural frequency of oscillation of node $i$, $\epsilon$ is the coupling strength, $A_{ij}$ represents the weight and $\phi_{ij}$ the phase-lag in the connection between node $i$ and $j$. To obtain the spatiotemporal dynamics of Kuramoto networks, we numerical integrate Eq.~\eqref{eq:km_model} using Euler's method, with timestep $dt = 0.001$.

Throughout this work, we consider different networks and parameters. For the globally connected case (all-to-all), $A_{ij} = 1, \, \forall \, i,j, \, \text{with} \, i\neq j$ and $A_{ij} = 0, \, \text{for} \, i = j$. For networks with nonlocal coupling, we consider k-ring graphs, where k is number of connections each node has on each side around the ring where the network is defined. In this case, $A_{ij} = 1$ if $d_{ij} \leq \mathrm{k}$, where $d_{ij} = \mathrm{min}(|i - j|, N - |i - j|)$, and $A_{ij} = 0$ otherwise. For distance-dependent coupling, $A_{ij} =\sfrac{1}{\eta(\alpha) (d_{ij})^\alpha}$, where $\eta(\alpha) = \sum_{j=1, j \neq i}^{N}d_{ij}^{-\alpha}$ with $\alpha$ being a parameter that controls the decay of the connection weights. The random network is obtained through Watts-Strogatz networks \cite{watts1998collective} with rewiring probability equal one. Lastly, the distance-dependent phase-delays are given by $\phi_{jk} = \sfrac{\pi d_{jk}}{\mathrm{k}}$.

\subsection{Order parameter -- measuring synchronization level}\label{sec:order_parameter}

To calculate the global level of phase synchronization in the simulations, we use the order parameter given by
\begin{equation}
    R(t) = \frac{1}{N} \left| \sum_{j=1}^{N} \exp{\Big(\i\theta_j(t)\Big)} \right|,
\label{eq:order_parameter}
\end{equation}
where $R(t)$ indicates the level of phase synchronization of the network at time $t$, with $R \in [0,1]$, where $R = 1$ indicates the network is phase synchronized, $R \sim \sfrac{1}{\sqrt{N}}$ is the expected value of an asynchronous state with random phases, and $R = 0$ is obtained in cases where there are symmetries in the phase relationship of nodes (anti-phase or a constant and nonzero phase offset).

\subsection{EEG dataset and analyses}\label{sec:eeg_data}

We analyze the EEG dataset from \cite{PhysioNet-eegmmidb-1.0.0}, which is an open, freely available dataset. Specifically, we consider the EGG data from 109 healthy human volunteers under two different resting state conditions. For each participant, two separate 1-minute EEG recordings were acquired at a sampling rate of 160 Hz: one with the eyes closed and another with the eyes open. Our goal is to distinguish between these two states based on the EEG signals. To do so, from the raw EEG signal, we filter the data in the $\alpha$-band (8 -- 12 Hz) and then obtain the phase associate with the signal in each channel in the band. To obtain the phases, we use the technique Generalize Phase, introduced in \cite{davis2020spontaneous}. With this, we obtain time series of phases for all 64 EEG channels for each participant under both resting-state conditions.

\section*{Code and data availability}

An open-source repository with codes used in this work is available at  \href{https://github.com/budzinskilab/phase-based-ordinal-analysis}{\textcolor{Cerulean}{github.com/budzinskilab}}. The EEG dataset analyzed in this study is freely available at \href{https://physionet.org/content/eegmmidb/1.0.0/}{\textcolor{Cerulean}{physionet.org}}.

\begin{acknowledgements}
We thank Cristina Masoller for the useful comments and discussion. R.J.S-S was supported by a MITACS GRA award and Coordena\c{c}\~ao de Aperfei\c{c}oamento de Pessoal de Nível Superior - Brasil (CAPES) - Finance Code 001. B.R.R.B acknowledges the suport of Conselho Nacional de Desenvolvimento Cient\'ifico e Tecnol\'ogico (CNPq) through fellowship and Grant No. 150376/2026-0 and INCT-NeuroComp. T.L.P acknowledges the financial support provided by Conselho Nacional de Desenvolvimento Cient\'ifico e Tecnol\'ogico (CNPq), Grants Nos. 305189/2022-0 and 408254/2022-0; the INCT-NeuroComp (CNPq Grant No. 408389/2024-9); Coordena\c c\~ao de Aperfei\c coamento de Pessoal de N\'ivel Superior (CAPES), Grants Nos. 88881.189166/2025-01. R.C.B acknowledges the support of the Natural Sciences and Engineering Research Council of Canada (NSERC) [grants RGPIN-2026-05758 and DGECR-2026-00066] and the support of the Canada Research Chairs program.
\end{acknowledgements}

\end{document}